\documentclass[twoside,british,3p]{elsarticle}
\usepackage[T1]{fontenc}
\usepackage[latin9]{inputenc}
\usepackage{xcolor}
\usepackage{textcomp}
\usepackage{amstext}
\usepackage{graphicx}
\usepackage{esint}
\PassOptionsToPackage{normalem}{ulem}
\usepackage{ulem}

\makeatletter

\newcommand{\lyxmathsym}[1]{\ifmmode\begingroup\def\b@ld{bold}
  \text{\ifx\math@version\b@ld\bfseries\fi#1}\endgroup\else#1\fi}

\providecolor{lyxadded}{rgb}{0,0,1}
\providecolor{lyxdeleted}{rgb}{1,0,0}

\journal{Cryogenics}

\renewcommand*{\l@figure}[2]{ %
\setlength\@tempdima{2.3em} %
\noindent\hspace*{1.5em}Figure #1\hfil\newline\newline }

\@ifundefined{showcaptionsetup}{}{%
 \PassOptionsToPackage{caption=false}{subfig}}
\usepackage{subfig}
\makeatother

\usepackage{babel}
\begin{document}

\title{A Compact Thermal Heat Switch for Cryogenic Space Applications Operating
near 100 K}

\author[rvt,focal]{M. Dietrich\corref{cor1}}

\ead{marc.dietrich@ap.physik.uni-giessen.de}

\author[rvt,focal]{A. Euler}

\author[rvt,focal]{G. Thummes}

\cortext[cor1]{Corresponding author. Address: Institute of Applied Physics, University
of Giessen, Germany. Tel.: +49 641 99 33462.}

\address[rvt]{Institute of Applied Physics, University of Giessen, D-35392 Giessen,
Germany}

\address[focal]{TransMIT-Center for Adaptive Cryotechnology and Sensors, D-35392
Giessen, Germany}
\begin{abstract}
A thermal heat switch has been developed intended for cryogenic space
applications operating around 100 K. The switch was designed to separate
two pulse tube cold heads that cool a common focal plane array. Two
cold heads are used for redundancy reasons, while the switch is used
to reduce the thermal heat loss of the stand-by cold head, thus limiting
the required input power, weight and dimensions of the cooler assembly.
After initial evaluation of possible switching technologies, a construction
based on the difference in the linear thermal expansion coefficients
(CTE) of different materials was chosen. A simple design is proposed
based on thermoplastics which have one of the highest CTE known permitting
a relative large gap width in the open state. Furthermore, the switch
requires no power neither during normal operation nor for switching.
This enhances reliability and allows for a simple mechanical design.
After a single switch was successfully built, a second double-switch
configuration was designed and tested. The long term performance of
the chosen thermoplastic (ultra-high molecular weight polyethylene)
under cryogenic load is also analysed.\end{abstract}
\maketitle

\section{Introduction}

In space technology, redundancy concepts are often used to minimize
the impact of a single failure. For applications requiring cryocoolers
this can be done in several ways \citep{RossJr.2002}. Nowadays cryocoolers
for space applications in the 77 K range are of the pulse-tube type
combined with a flexure bearing compressor. Possible sources of failure
in such a configuration are the control electronics and the compressor
(e.g. coils, spring breakage, seals), and the pulse-tube cold head
(e.g. gas leakage). Additionally, the cold head can suffer from a
long term gas contamination by outgassing components in the compressor
or cold head itself.

As part of a research project studying various redundancy concepts
for satellite operations, a heat switch was developed including two
cold heads (CH1 and CH2) mounted on a single focal plane array detector
\begin{figure}
\centering{}\includegraphics[width=17cm]{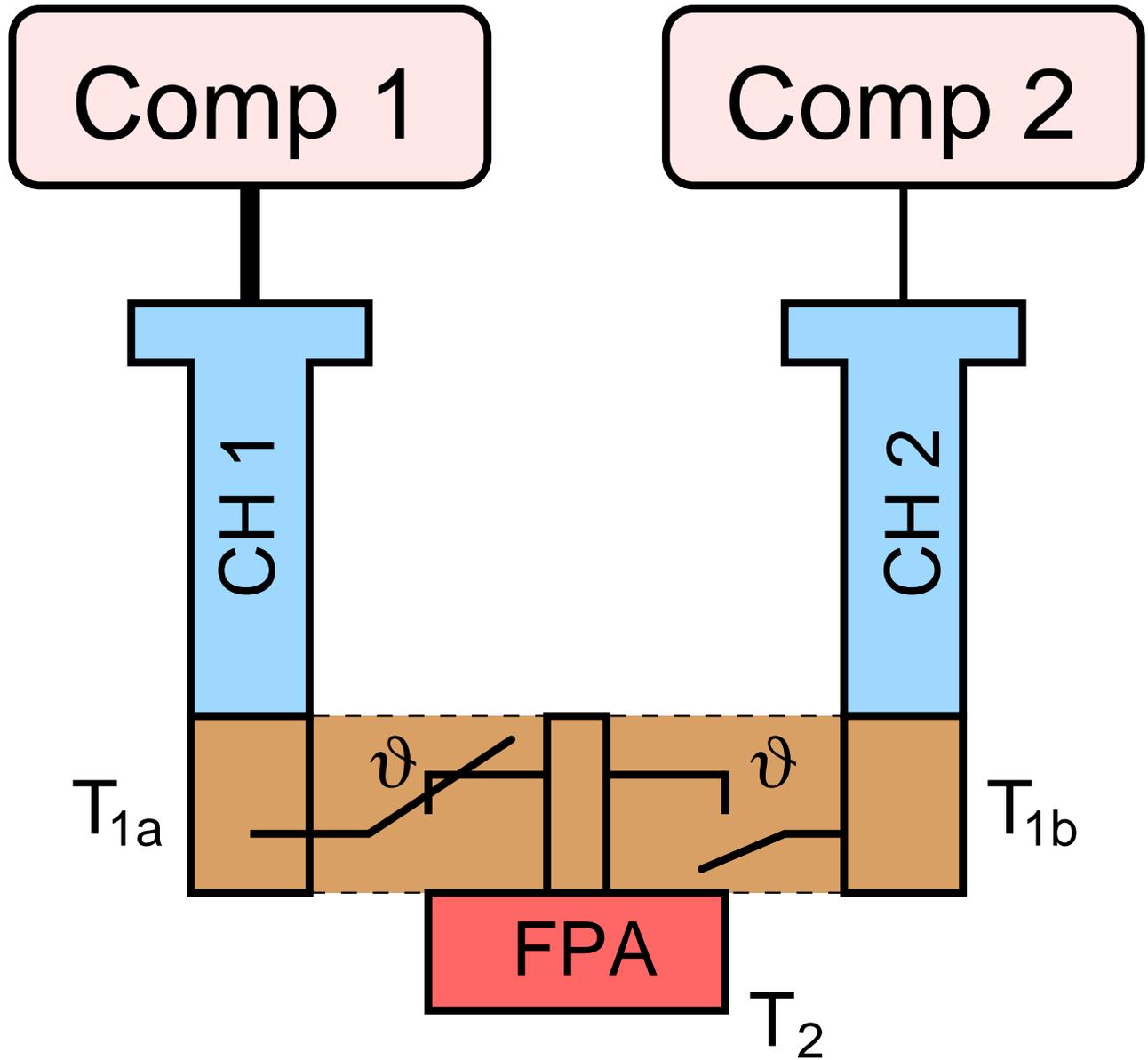}\caption{Schematic of the cold head assembly including heat switch and FPA.\label{fig:cooler_assembly}}
\end{figure}
(FPA, see figure \ref{fig:cooler_assembly}). The heat switch thermally
connects an active cold head to the FPA while increasing the thermal
contact resistance $\vartheta$ to a second cold head in stand-by
mode, thus reducing the heat load of the stand-by system to the active
cold head. In case of a failure of the active cold head, the stand-by
cryocooler is turned on and the heat switch thermally disconnects
the malfunctioning system while thermally connecting the now active
redundant cryocooler. The switching is done automatically without
need for external control or power.

There exist various kinds of heat switches for space applications,
each having their own advantages and disadvantages. Most common are
heat switches of the Gas-Gap and CTE-based (CTE: linear thermal coefficient
of expansion) type \citep{Chan1990,Marland2004,Wang2007}, but also
some designs based on other physical effects are known \citep{Prenger1994,You2005}.
In Gas-Gap switches the pressure of a gas in a small gap is controlled.
This can be done e.g. by use of adsorbers or valves connected to gas
reservoirs. The presence/absence of the gas in the small gap (typically
$\ll$ 100 \textmu{}m) enables/disables a thermal contact. The use
of adsorbers is preferred for space applications as it enables passive
switching. The design is somehow complicated because it involves the
choice of a proper combination of gas sort, filling pressure, and
adsorption material for a given temperature range. Especially for
relatively high temperatures this can be challenging \citep{Catarino2008}.
On the other side, this switch type does not contain any moving parts
which may fail during the lifetim\textcolor{black}{e, e.g. because
of wear or mechanical breakage.}

The CTE-based switches rely on the thermal expansion of one or more
components. In the most common mode of operation, a small gap separates
two solids one of which has a high CTE compared to the other one.
When the temperature decreases the high CTE material ``shrinks''
and closes the gap between the two solids. Below a certain ``switching
temperature'', the gap is fully closed, thus providing a heat conduction
path. Further decrease in temperature increases the contact pressure
and therefore lowers the thermal contact resistance. While the CTE-based
switch is less complex compared to the gas-gap variant, it shares
with it the disadvantage of requiring a relatively small gap in the
order of several micrometers. However, by using thermoplastics with
a high CTE, such as ultra-high molecular weight polyethylene (UHMW-PE),
in order to circumvent the requirement of tiny gaps, the CTE-based
switch becomes superior compared to other designs with respect to
the required temperature range and standards for space applications.
The aim of this work was to provide a proof of concept; so properties
like minimum weight, high stability, and device integration, which
are required for space applications, had a minor priority.

\section{Thermal heat switch concept}

Based on the design considerations outlined in the previous section,
a CTE-based switch was chosen. Because the switch also needs to serve
as a support for the FPA, the contact areas that are connected to
the cold head and FPA must not move upon switching. As another requirement,
the switch should exhibit an on-state thermal conductivity of more
than 1 W/K at an operating temperature between 80-100 K and an off-state
conductivity of less than 1 mW/K.

\begin{figure}
\begin{centering}
\includegraphics[width=17cm]{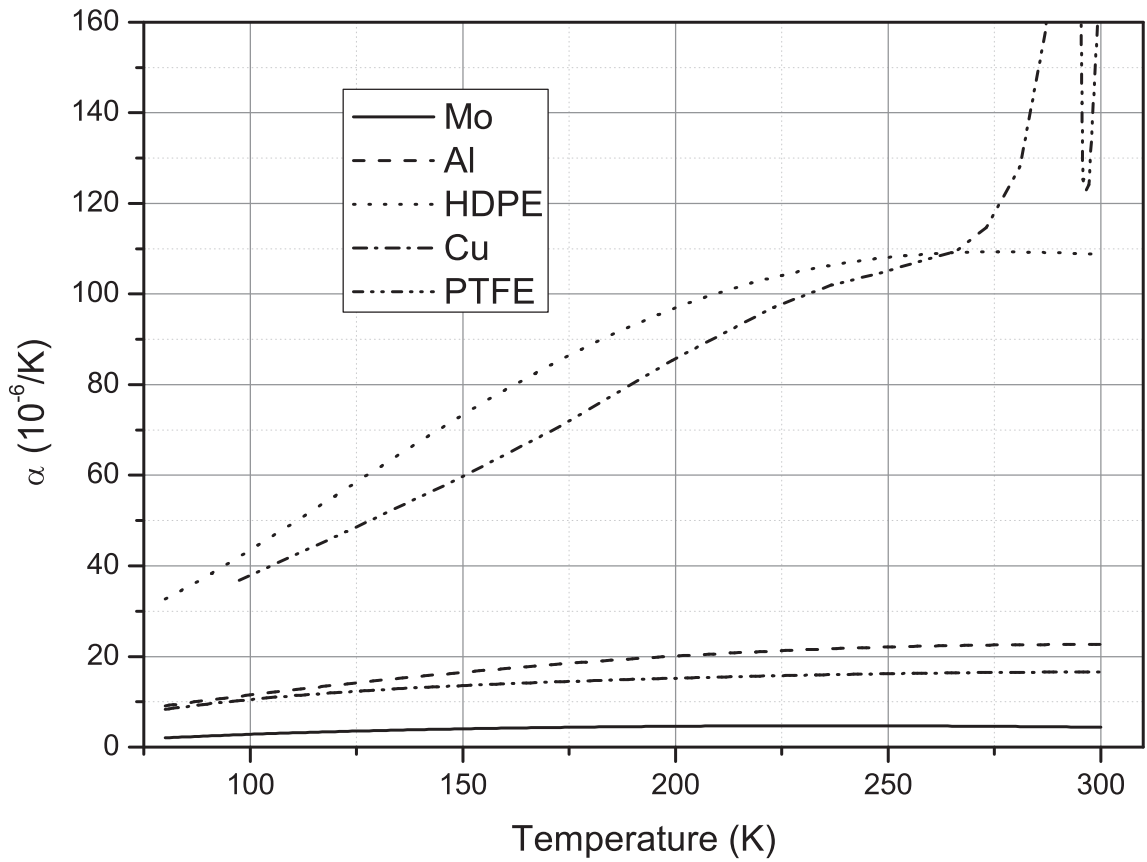}
\par\end{centering}

\caption{Coefficient of thermal expansion (CTE) versus temperature for Copper,
Aluminium, Molybdaenum and PTFE. Data for metals were taken from the
NIST database \citep{Nist2012}, except for HDPE which was taken from
Hartwig \citep{Hartwig1994} and PTFE which was taken from \citet{Kirby1956}.
\label{fig:CTE}}
\end{figure}

In many switches, the high-CTE component itself also acts as a thermal
conductor. Unfortunately, high thermal conductivity materials, such
as metals, exhibit a rather small CTE leading to a small gap sizes
which require careful manufacturing and increase the risk of failure.
Thermoplastics on the other hand, have a relatively high CTE compared
to metals but a low thermal conductivity. This leads to a design where
a thermoplastic is used as the switching element only, bringing two
metals with high thermal conductivity into contact. Figure \ref{fig:CTE}
 shows the CTE between 80 K and 300 K of some metals and thermoplastics;
the data for metals were taken from the NIST database \citep{Nist2012}.
While most metals exhibit a CTE of 10-20$\times10^{-6}/K$, polytetrafluoroethylene
(PTFE) for example has an order of magnitude higher CTE compared to
copper at room temperature, but as for all polymers this strongly
depends on the composition. PTFE shows two solid-solid phase transitions
near room temperature with a maximum CTE of more than 500$\times10^{-6}/K$
\citep{Kirby1956}. Our measurements using liquid nitrogen revealed
that ultra-high-molecular-weight polyethylene (UHMW-PE) has an even
higher thermal contraction than PTFE between room temperature and
77 K, though we couldn't find any CTE data for UHMW-PE at cryogenic
temperatures. So we initially based our calculations on the HDPE (high-density
polyethylene) CTE data from \citep{Hartwig1994}.

\begin{figure}
\begin{centering}
\subfloat[Schematic sectional drawing\label{fig:single_switch_a}]{\includegraphics[width=8cm]{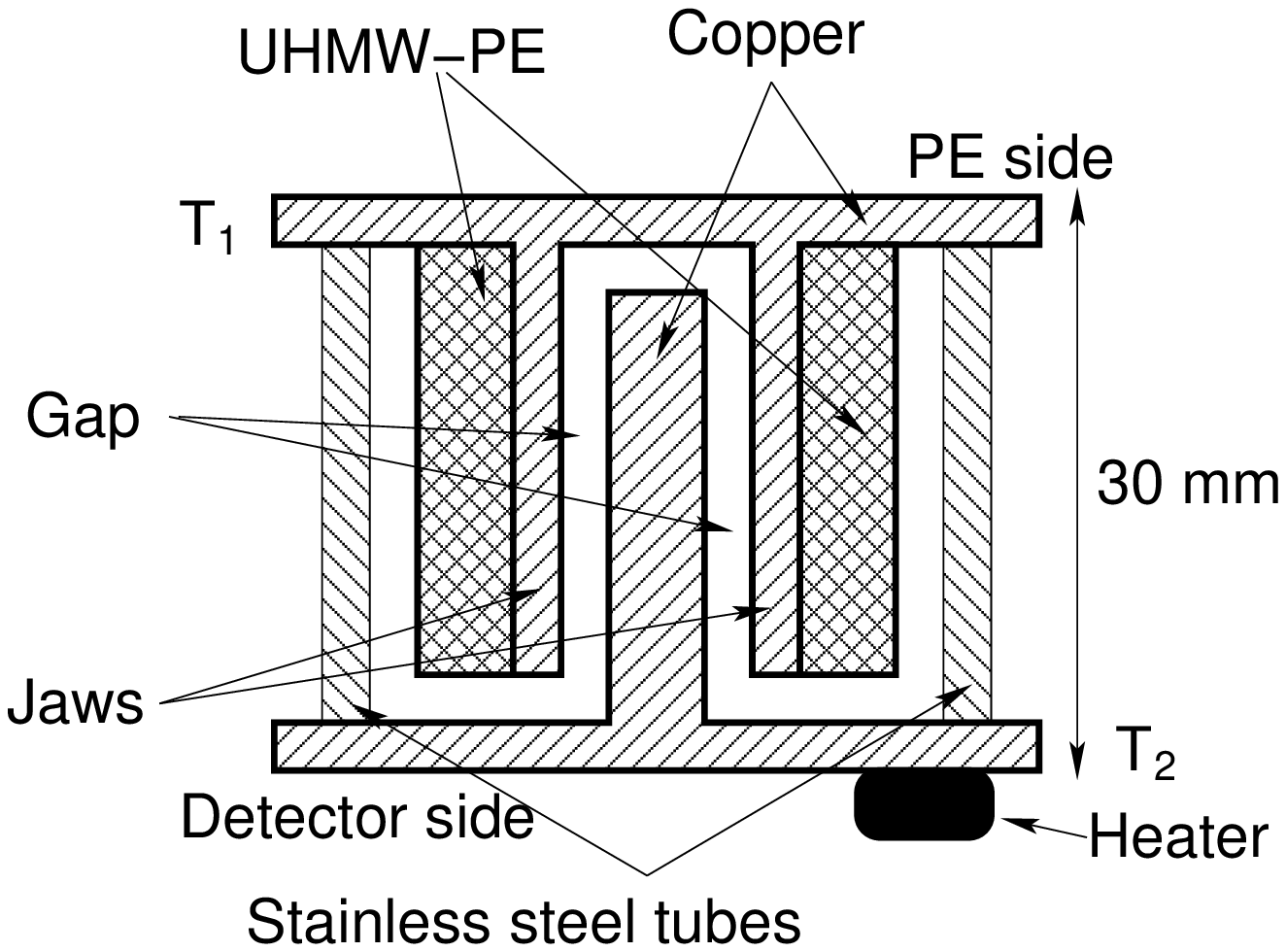}}\subfloat[3D-Model (the gap is magnified for clarity)\label{fig:single_switch_b}]{\includegraphics[width=8cm]{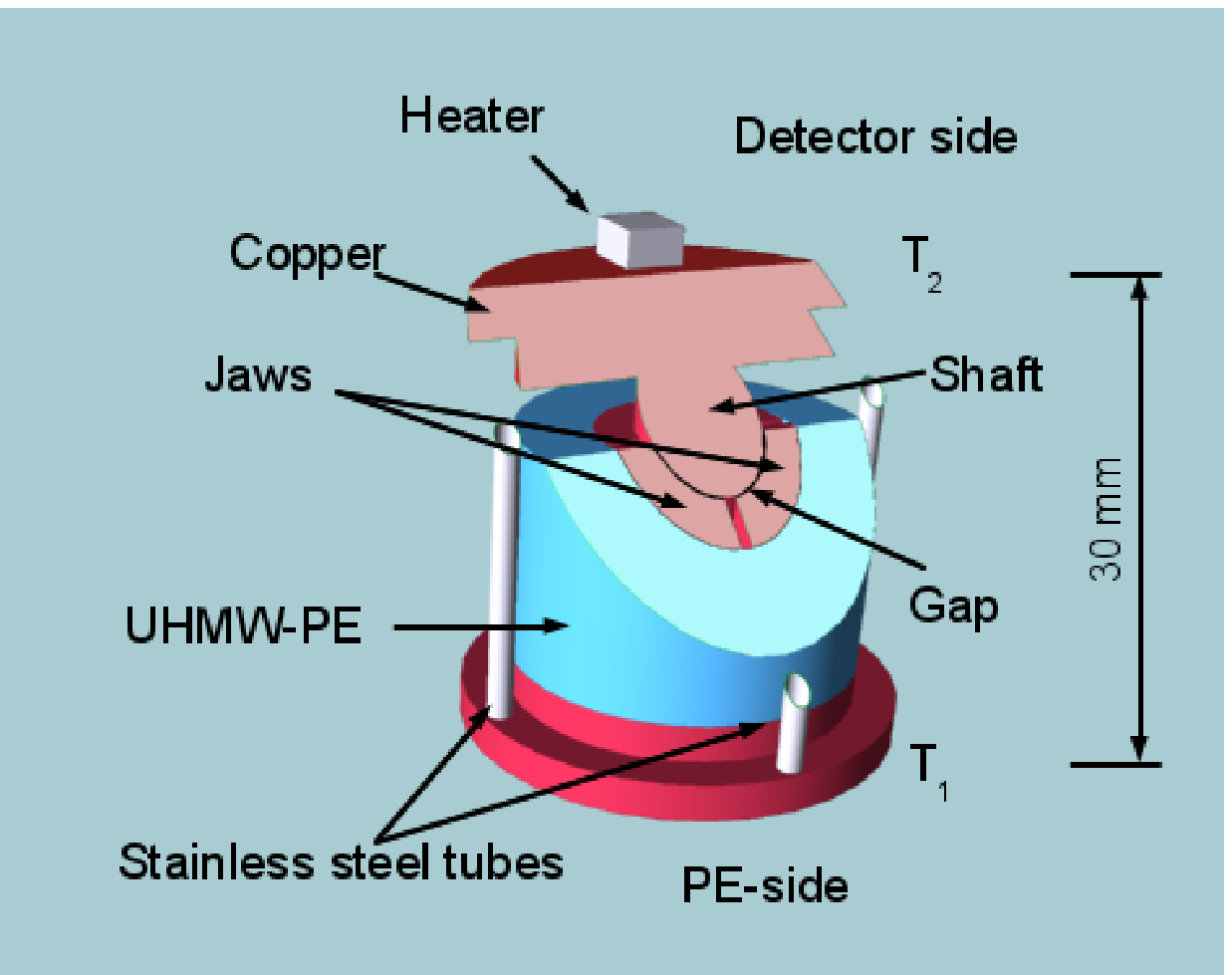}}
\par\end{centering}

\caption{Design of the single switch.\label{fig:single_switch} }
\end{figure}

When designing a CTE-based switch, one important parameter is the
gap width, which depends on the CTE material and the desired switching
temperature. One has to take into account, that the standy-by cold
head will cool down because of the finite off-state conductance. In
our case, an off-state conductance of 1 mW/K would cool the stand-by
cold head down to about 220 K. This means, that the switch needs to
change state below this temperature. Otherwise the switch would also
close on the stand-by side producing a thermal short. On the other
hand, a high on-state switching temperature is desired to achieve
a high contact pressure and thus a low thermal resistance. For HDPE,
a CTE of

\[
\alpha(T)=28.9-0.6338*T+0.01178*T^{2}-4.51254*10^{-5}*T^{3}+5.28582*10^{-8}*T^{4}
\]

as extracted from the graph in Appendix 4A of \citep{Hartwig1994}
was used for calculations. The contraction in radial direction of
the cylindrical switch (see figure \ref{fig:single_switch_b}) is
calculated by
\[
\Delta r=R_{0}\intop_{Tw}^{Tc}\alpha(T)\, dT,
\]

where $T_{c}$ is the switching temperature and $T_{w}$ is the warm
temperature at which the gap size is measured. $R_{0}$ is the inner
radius of the UHMW-PE cylinder at room temperature. The copper shaft
and jaws also shrink, but their shrinking was ignored in the initial
gap calculations, since it is about an order of magnitude less than
that of the UHMW-PE. Based on the considerations above, a gap width
of 80 micrometers at room temperature $T_{w}$ was chosen. The remaining
50 microns which the UHMW-PE would additionally shrink from $T_{c}=$200
K to the detector operating temperature of 100 K, if there will be
no shaft, are turned into increased contact pressure. During the initial
testing, the gap was adjusted several times to account for the higher
CTE of UHMW-PE and the thermal expansion of the copper shaft and jaws.

Two switches were built: a single switch connected to a single cold
head and heat load, which was used for initial testing. Later on,
a second switch was built which has a T-form to connect two cold heads
to a single load located in the middle of the switch. Figure 
\ref{fig:single_switch_a} shows the sectional drawing and Figure
\ref{fig:single_switch_b} shows a 3D-model of the single, cylindrical
switch design. The part connected to the heat load (detector side)
consists of an inner shaft made of a solid copper cylinder (10 mm
diameter) with a flange on one end. The part connected to the cold
head (PE-side) consists of a copper flange with four integrated copper
jaws that are separated from the inner cylinder by the gap. A non-enforced
hollow UHMW-PE (virgin Tivar$^{\lyxmathsym{\textregistered}}$1000
from Quadrant PHS GmbH, Vreden, Germany) cylinder is put around the
jaws to act as the high-CTE element. The two copper parts are hold
together by four thin stainless steel tubes ($\textrm{Ø}$ 2 mm, 150
\textmu{}m wall thickness) which mainly determine the thermal off-state
resistance. The total height of the switch is 30 mm.

The distribution of stress inside the switch components was calculated
using Hooke's law which can be found in textbooks. The contact pressure
of the jaws to the shaft at 100 K was estimated to be 1.4 MPa, while
the maximum tensile stress in the UHMW-PE was estimated to be 5 MPa.
Figure 
\begin{figure}
\begin{centering}
\includegraphics[width=17cm]{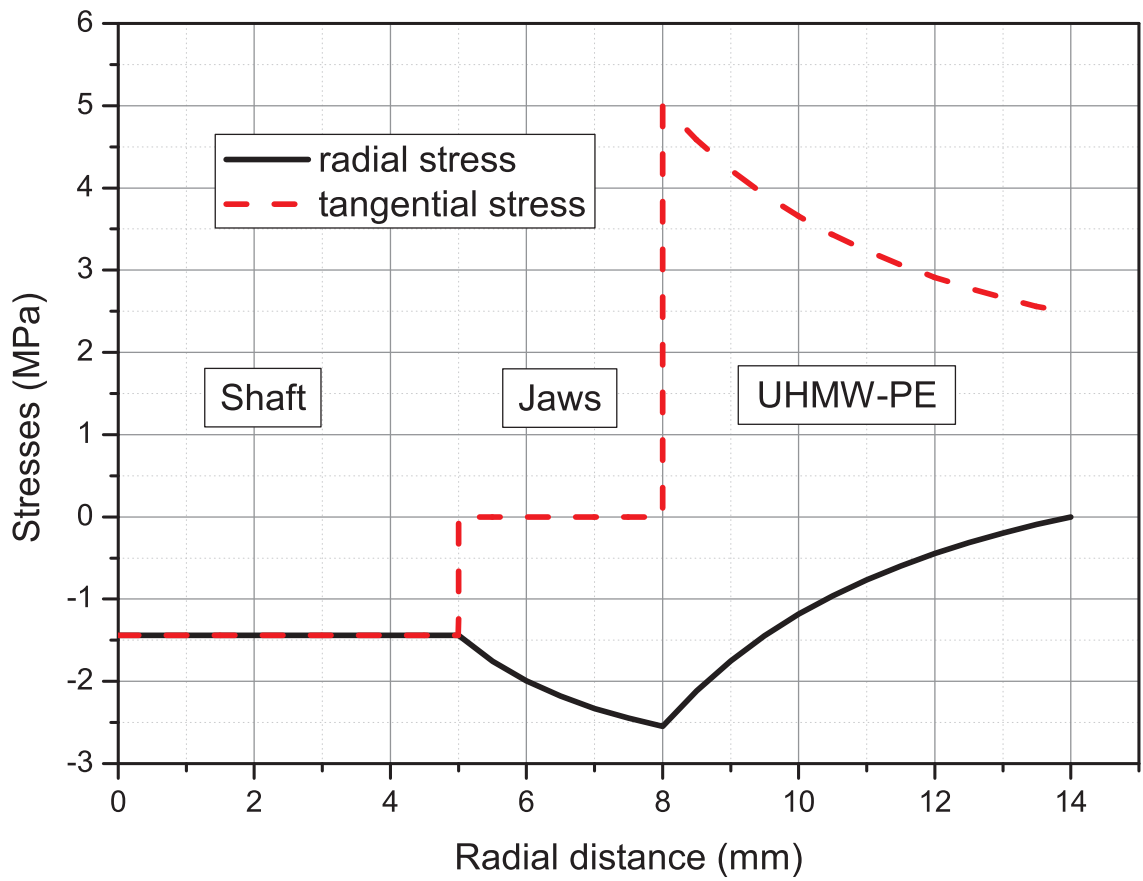}
\par\end{centering}

\caption{\label{fig:stress}Radial- and tangential (circumferential) stresses
inside the switch components at 100 K.}
\end{figure}
\ref{fig:stress} shows the compressive and tensile stresses in radial
and tangential (= circumferential) directions inside the switch components,
calculated at an operating temperature of 100 K. The axial stresses
were omitted in the calculation.

\section{Performance testing}

The test apparatus consists of a coaxial pulse tube cold head driven
by an AIM SL400 linear compressor \citep{Yang2005} to which the single
switch is attached. The detector side of the switch is equipped with
an electrical heater. Pt100 temperature sensors are placed at the
PE and the detector side of the switch, named $T_{1}$ and $T_{2}$
respectively (see figure \ref{fig:single_switch_b}). For radiation
shielding five layers of superinsulating foil are wound around the
cold head and the switch.

The single switch was tested in several cool down/heat up cycles.
Figure 
\begin{figure}
\centering{}\includegraphics[width=17cm]{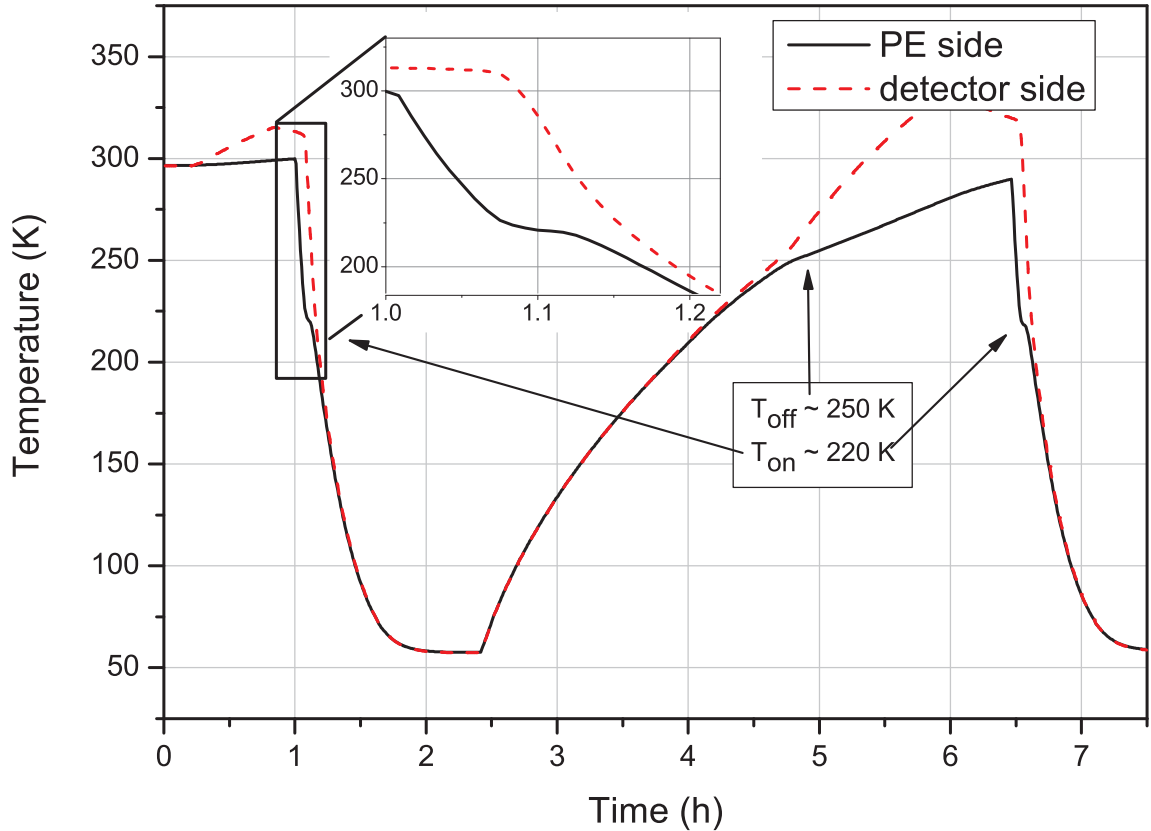}\caption{One cycle of the single switch in actuation.\label{fig:single_switch_actuation}}
\end{figure}
\ref{fig:single_switch_actuation} shows the switch in a single actuation
cycle. During the measurement a constant heat load of 500 mW is being
applied to the detector side of the switch. The inset in Figure \ref{fig:single_switch_actuation}
shows the cool down process to the closing temperature in detail.
The detector side of the switch maintains a constant temperature until
the switch closes at a temperature of about 220 K. After that, the
detector side cools down quickly until it reaches the cold head temperature.
From there on, both temperatures further decrease until the base temperature
of 57 K is reached. At 2.5 hours the compressor is turned off. Both
temperatures start to rise until the switch opening temperature of
about 250 K is reached. From there on, the cold head temperature rises
slower than the sensor temperature because of the low thermal coupling
in the off-state. After 6 hours the heater is switched off and after
around 6.5 hours the cooler and heater are started again and the next
cycle begins. There exists a small hysteresis of about 25 K due to
the lag of the UHMW-PE temperature with respect to the cold head temperature.

The close and open temperatures can be adjusted by the gap width between
the copper shaft and the copper jaws. Increasing the gap by reducing
the shaft diameter reduces the switching temperatures. Figure 
\begin{figure}
\centering{}\includegraphics[width=17cm]{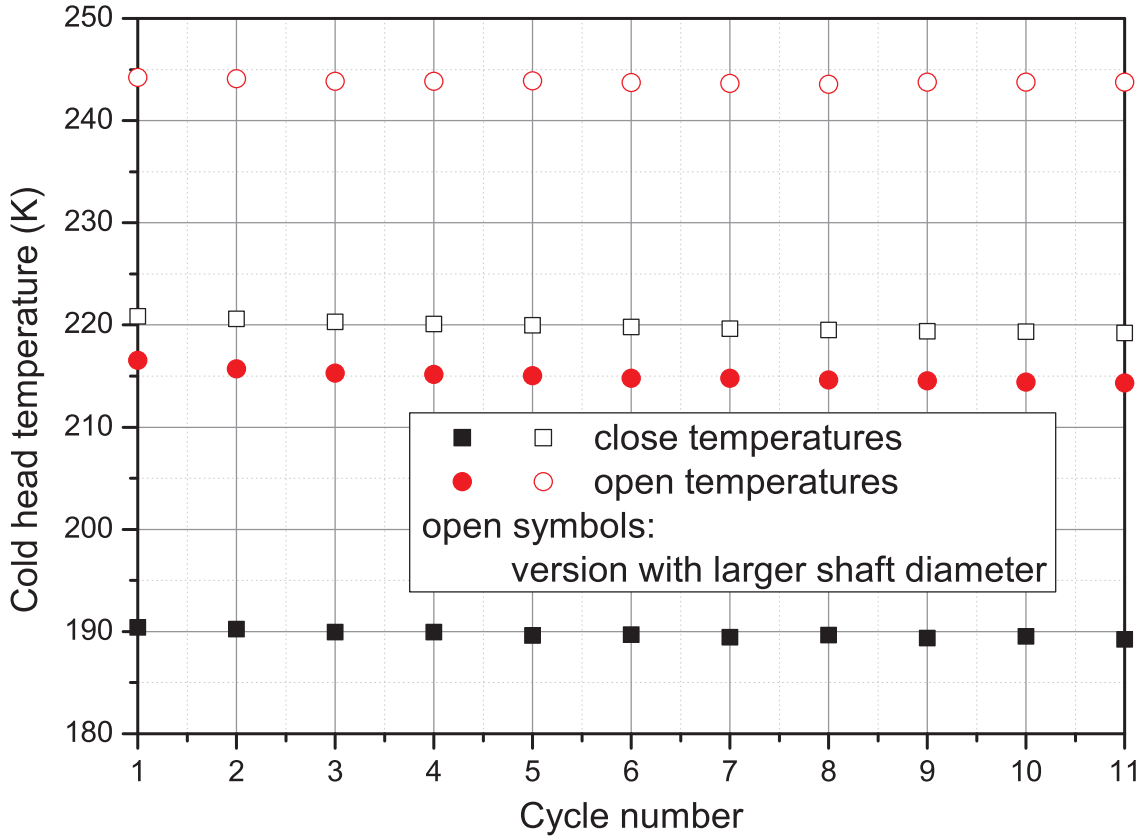}\caption{Open/Close temperatures measured over several switch cycles.\label{fig:zyklen}}
\end{figure}
\ref{fig:zyklen} illustrates the close and open temperatures of the
switch for 11 cool down/heat up cycles and for two diameters of the
shaft. The open symbols show the close/open temperatures for the switch
with larger shaft diameter, where the gap was $\sim$ 20 \textmu{}m
smaller than that of the switch with smaller shaft diameter (solid
symbols). Over the 11 open/close cycles, the switch temperatures are
almost equal. If looking closely to the graph, a small tendency to
lower temperatures with increasing cycle number can be seen, which
can be attributed to creep of the UHMW-PE material (see section \ref{sec:Long-term-performance}). 

\begin{figure}
\centering{}\includegraphics[width=17cm]{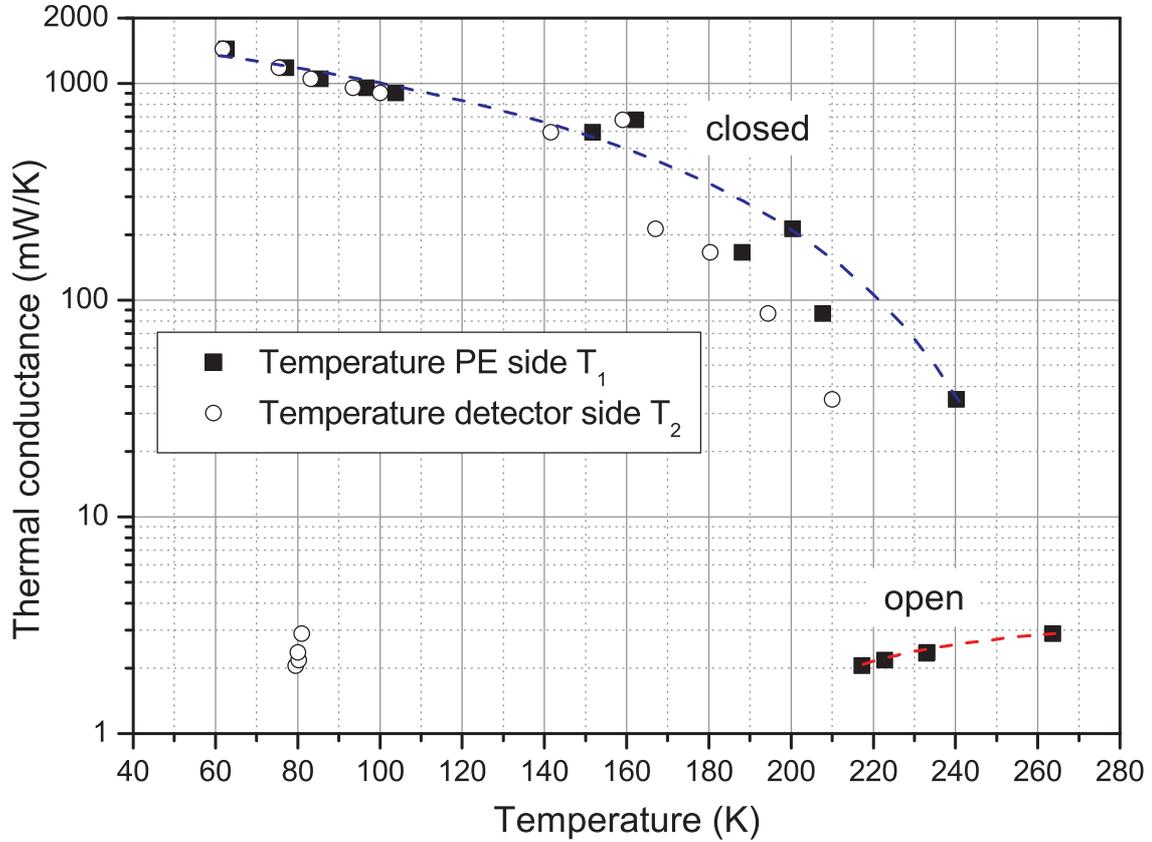}\caption{Heat conductance of the single switch in on- and off-state as function
of temperature. The difference between corresponding $T_{1}$and $T_{2}$
data points represents the temperature difference between the two
sides of the switch. The lines serve as guide to the eyes.\label{fig:heat_conductivity}}
\end{figure}

In order to measure the thermal conductance of the single switch,
it was mounted in opposite direction with the detector side ($T_{2})$
connected to the cold head while the heater was mounted at the PE-side
($T_{1}$) of the switch. In this way it is possible to also measure
the off-state conductance at different stand-by cold head temperatures.
During the on-state measurements, a constant heat load of 1 W is applied
and the input power to the cold head is controlled so that the desired
temperature is reached. The only exception is the point at $T_{1}$=
200 K where the heat load was 7 W. The conductance is given by the
ratio of the applied heat load $\dot{Q}_{c}$ and the temperature
difference between the two sides of the switch ($T_{1}-T_{2}$). For
measuring the off-state conductance, the heater power $\dot{Q}_{c}$
is tuned in such a way that the switch is kept open while the detector
side is cooled down to 80 K. Figure~\ref{fig:heat_conductivity}
shows the off-state conductance as function of temperature. The heating
power was varied between 130 and 180 mW to keep the PE-side of the
switch at the desired temperature. At a typical off-state operation
point of 220 K, the conductance is around 2 mW/K. For higher temperatures,
the off-state conductance increases because of the higher heat conductance
of the stainless steel support tubes. For on-state measurements, the
heater is switched off until the switch closes due to heat conduction
losses over the stainless steel tubes. Once the switch is closed,
the heater can be used to measure the thermal conductance. Figure
\ref{fig:heat_conductivity} also shows the on-state conductance in
a temperature range between 80 and 240 K. Due to the higher contact
pressure at lower temperatures, the thermal conductance increases
with decreasing temperature and consequently the temperature difference
$T_{1}-T_{2}$ becomes smaller. For an operation point of 100 K, the
thermal conductance in on-state is 1000 mW/K. 

To test the mechanical stability, the switch with the smaller shaft
diameter has undergone a shaker test at AIM GmbH (Heilbronn), where
the switch was shaked at different frequencies with a 150 g mass mounted
on the detector side. The test revealed several resonance points at
which the two switch parts (jaws and shaft) clashed together. An optical
inspection after the test confirmed this, but no hints of deformation
could be seen. Figure 
\begin{figure}
\centering{}\includegraphics[width=17cm]{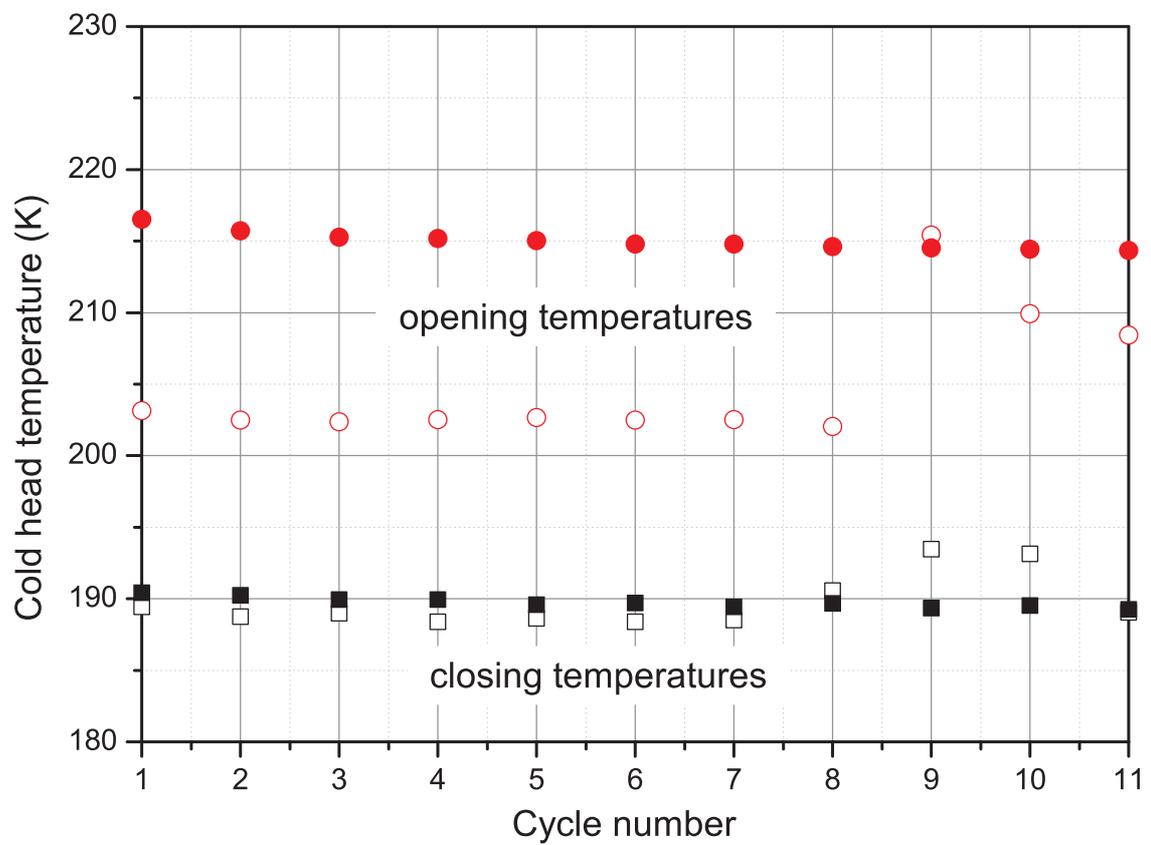}\caption{Open/close temperatures before (solid symbols) and after shaker test
(open symbols); switch version with smaller shaft diameter.\label{fig:o/c_after_shake}}
\end{figure}
\ref{fig:o/c_after_shake} shows the open/close temperatures after
the shaker test. Compared to the cyclic test in Figure \ref{fig:zyklen}
(solid symbols), the opening temperature dropped by about 13 K, while
the closing temperature is nearly the same as before. In the measurements
after the shaker tests, between cycle 8 and 9 the switch was heated
up to 340 K, causing an unusual shift to higher open/close temperatures.
The root cause of this is not yet understood. The thermal conductance
measurements in Figure 
\begin{figure}
\centering{}\includegraphics[width=17cm]{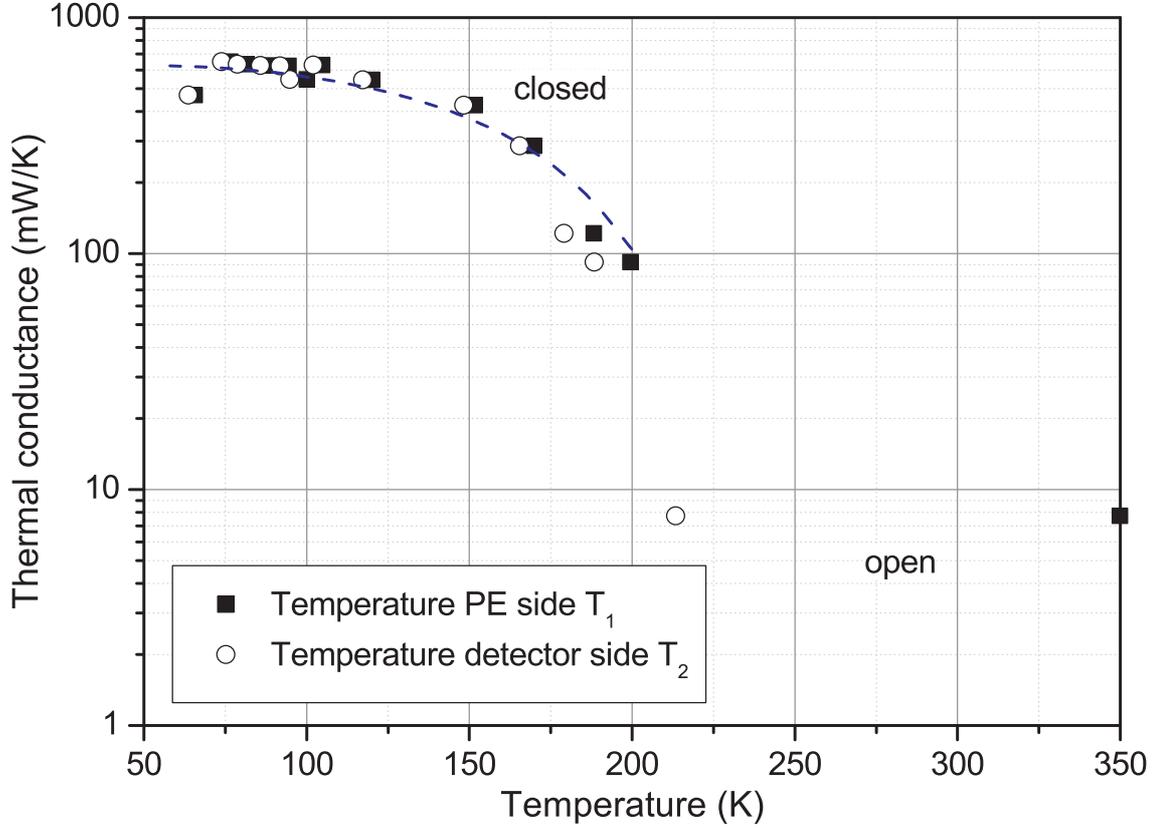}\caption{Heat conductance in on- and off-state after the shaker tests. The
difference between corresponding $T_{1}$ and $T_{2}$ data points
represents the temperature difference between the two sides of the
switch. The lines serve as guide to the eyes. \label{fig:heat_cond_after_shaker}}
\end{figure}
\ref{fig:heat_cond_after_shaker} show a decrease in on-state conductance
to $\sim$ 600 mW/K at 100 K, while the off-state conductance increased
to $\sim$ 8 mW/K after the shaker test. All measurements were done
with 1 W of heating power. These results hint at a small deformation
of the switch by the shaker test, which likely affects the heat conductance.
Given that the switch was not yet optimized for mechanical stability,
the results after the shaker test show that the present design using
a large gap, compared to that of purely metallic CTE-based switches,
can simplify heat switch manufacturing while maintaining a reliable
switching function.

After successful tests with the single switch, we built a T-formed,
double switch, which basically consists of two singles switches with
a common detector side. The test apparatus was extended to include
two pulse tube cold heads of the same type connected to the two PE-sides
of the T-form switch. A heater and a temperature sensor ($T_{2}$)
were installed at the common detector side, and two temperature sensors
($T_{1A}$, $T_{1B}$) and two heaters were mounted on the PE-sides
of the double switch.

\begin{figure}
\centering{}\includegraphics[width=17cm]{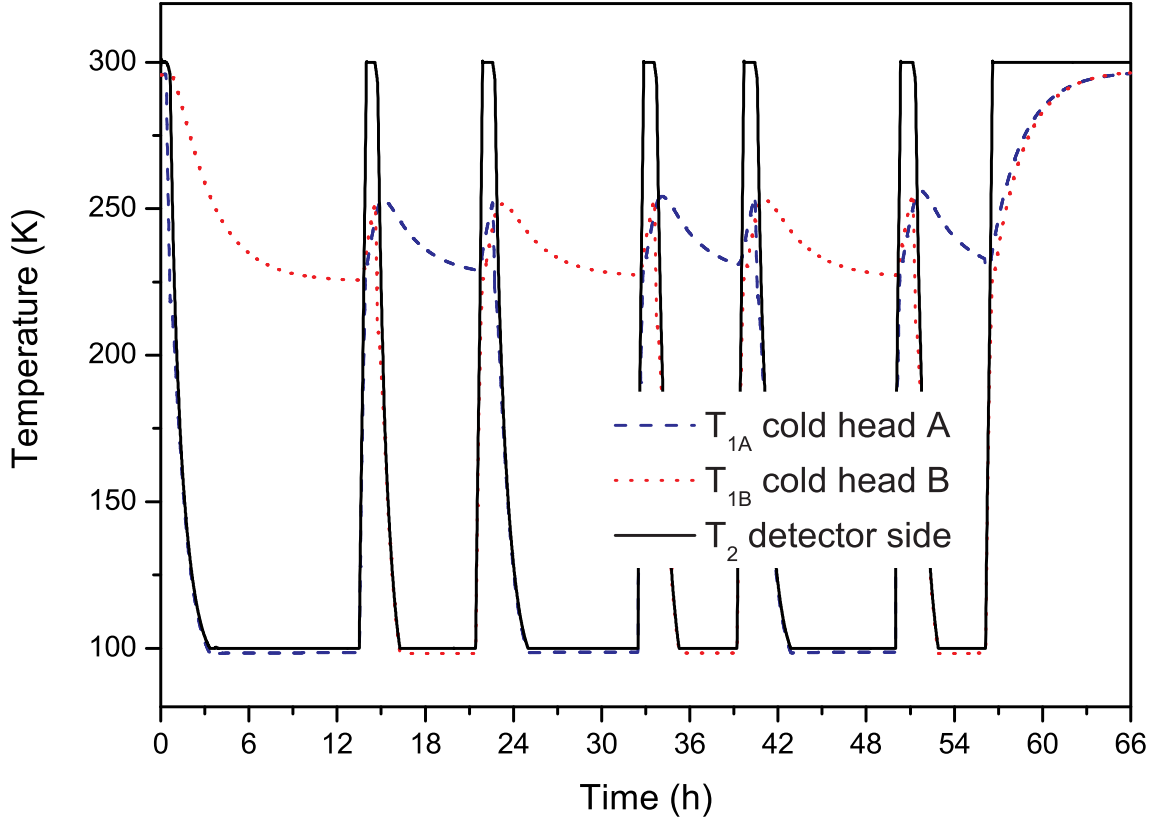}\caption{Cooldown/heat up cycle of the double switch.\label{fig:double_switch}}
\end{figure}

The cycle test results of the double switch are shown in Figure \ref{fig:double_switch}.
One of the coolers (cold head A) was started while the other was in
stand-by. After reaching a temperature of $T_{2}$ = 100 K, the heater
at the detector side was adjusted to maintain this temperature. The
stand-by side also cools down a bit because of the small but non-zero
off-state conductance of the switch. In general, the switch temperature
in off-state is given by a balance between heat conduction along the
switch and along the standby cold head. After the stand-by cold head
B reached a stable temperature of $T_{1B}$ \ensuremath{\approx} 225
K, the active cold head A was turned off and the cold-head side of
the switch was heated above the opening temperature again. Now the
active and the stand-by cold head changed their function, and the
cycle was repeated several times, as seen from Figure 
\ref{fig:double_switch}. Both sides of the switch are working as
expected. The opening/close temperatures are nearly identical. The
passive PE-side cools down to 225 K, which is only 25 K above the
closing temperature of the switch. By further lowering the thermal
conductance of the switch in off-state or reducing the heat leak through
the stand-by cooler this temperature difference can be enlarged in
future, as in the present lay-out it seems a bit low for long term
operation.

\section{Long term performance of UHMW-PE\label{sec:Long-term-performance}}

For long term satellite missions it is essential to know how the materials
used in the switch will degrade in their properties during mission
lifetime. It is known that thermoplastics tend to creep over time,
which would have a significant effect on the switch performance. Creep
behaviour is influenced by many aspects of the actual material composition
and treatment, so the results shown below should be handled with care.
Material pre-aging can significantly reduce long term creep, whereas
other methods like material-enforcement tend to reduce the CTE. 

Long term creep measurements for UHMW-PE have been studied for room
temperatures and above, mainly because of their application in the
medical sector \citep{Deng1998}. But creep data at cryogenic temperature
ranges is not available in literature. We therefore built a test apparatus
for testing the creep of our UHMW-PE samples at low temperatures using
strain gauges (type Micro-Measurements EK-13-250BF-10C/W). The UHMW-PE
samples were in form of a solid cube with an edge length of 20 mm.
Figure 
\begin{figure}
\centering{}\includegraphics[width=17cm]{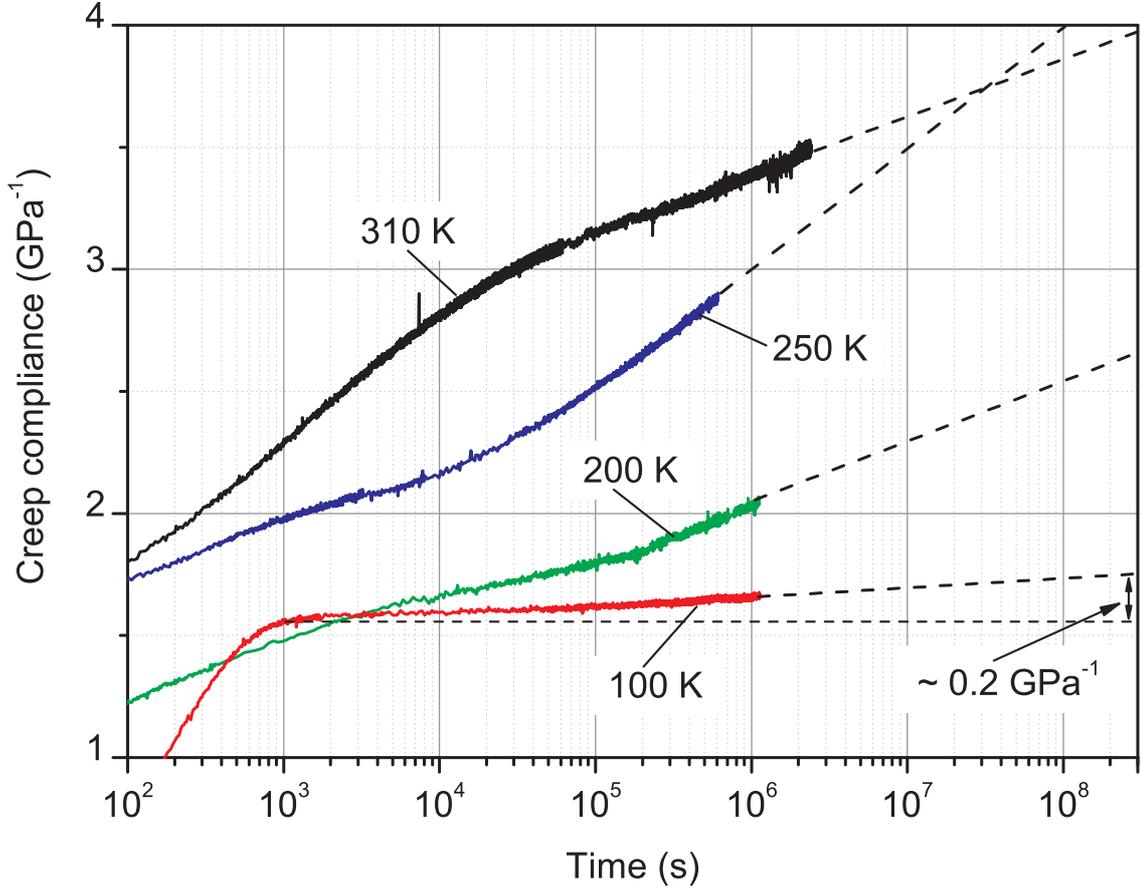}\caption{Measurement and prediction of long term creep of UHMW-PE at different
temperatures.\label{fig:creep}}
\end{figure}
\ref{fig:creep} shows the compliance data under a compressive load
of 1 MPa for several temperatures. Similar results are expected for
tensile stresses. The compliance $D$ is defined as $D(t):=\epsilon(t)/\sigma$,
where $\epsilon(t)$ is the measured strain and $\sigma$ is the applied
pressure load. After some initial relaxation processes on a time scale
of less than 10,000 s, it appears that the material starts to creep
linearly on a logarithmic time scale. At least for the 100 K data
this is in accordance with the work of Struik \citep{Struik1989},
who used short term measurements near room temperature to predict
low temperature creep data for temperatures $T<T_{g}$, where $T_{g}=130\, K$
is the glass transition temperature of UHMW-PE. For a mission time
of 10 years, a creep compliance of $0.2\, GPa^{-1}$ at 100 K can
be roughly extrapolated. For a typical maximum internal stress of
about 5 MPa (see figure \ref{fig:stress}), this would result in a
strain of only 0.1\%. When applied to our switch geometry, this corresponds
to a decrease in contact pressure by about 12\%.

\section{Conclusions }

A simple, compact, high reliability thermal heat switch for cryogenic
space applications operating near 100 K based on the thermal expansion
was built and tested. Two variants have been studied: a single and
a double heat switch configuration. The single switch showed a state
change around 220 K, and an on/off-state conductivity of more than
1 W/K and an 3 mW/K respectively. After shaker tests the performance
decreased a bit, but the switching function was not affected.

The double switch was successfully tested in a two cooler configuration
and showed reliable switching characteristics over several cycles.
The switching temperatures as well as the on/off state conductivity
was similar to the single switch design.

UHMW-PE, which was used as the high CTE material, shows a rather high
creep rate under uniaxial pressure at room temperature. To estimate
the degeneration of the material during switch operation at cryogenic
temperatures, creep tests were performed and extrapolated for long
term prediction. At 100 K, the compliance is estimated to be $0.2\, GPa^{-1}$
in 10 years resulting in a 12\% drop in contact pressure during on-state.

The CTE-based thermal switch presented in this paper is a promising
concept. Further development will focus on mechanical properties as
stability and weight. The long term creep of the UHMW-PE CTE material
also needs a more thorough investigation.

\section*{Acknowledgements}

This work was financially supported by the German Federal Ministry
of Economics and Technology (grant no. 50EE0940). The authors thank
AIM Infrared Modules (Heilbronn) for performing the shaker tests and
Frank Schmülling (DLR Bonn) for useful discussions.\newpage{}

\bibliographystyle{unsrtnat}
\bibliography{heat_switch}

\end{document}